# Diffractive Imaging of Coherent Nuclear Motion in Isolated Molecules

Jie Yang[1], Markus Guehr[2,3], Xiaozhe Shen[4], Renkai Li[4], Theodore Vecchione[4], Ryan Coffee[4], Jeff Corbett[4], Alan Fry[4], Nick Hartmann[4], Carsten Hast[4], Kareem Hegazy[4], Keith Jobe[4], Igor Makasyuk[4], Joseph Robinson[4], Matthew S. Robinson[1], Sharon Vetter[4], Stephen Weathersby[4], Charles Yoneda[4], Xijie Wang[4], Martin Centurion[1]

[1]University of Nebraska-Lincoln, 855 N 16th Street, Lincoln, Nebraska 68588, USA

[2]PULSE, SLAC National Accelerator Laboratory, Menlo Park, CA 94025, USA

[3]Physics and Astronomy, Potsdam University, 14476 Potsdam, Germany

[4]SLAC National Accelerator Laboratory, Menlo Park, CA 94025, USA

Corresponding authors

Markus Guehr, mguehr@uni-potsdam.de

Xijie Wang, wangxj@slac.stanford.edu

Martin Centurion, martin.centurion@unl.edu

**Abstract**. Observing the motion of the nuclear wavepackets during a molecular reaction, in both space and time, is crucial for understanding and controlling the outcome of photoinduced chemical reactions. We have imaged the motion of a vibrational wavepacket in isolated iodine molecules using ultrafast electron diffraction with relativistic electrons. The time-varying interatomic distance was measured with a precision 0.07 Å and temporal resolution of 230 fs full-width at half-maximum (FWHM). The method is not only sensitive to the position but also the shape of the nuclear wavepacket.

Photo-induced reactions are of particular interest for understanding the fundamental mechanisms driving the conversion of light into chemical and kinetic energy on ultrafast time scales. The coherent nuclear motion is particularly important to study the reaction pathway and energy conversion efficiency in processes that cannot be described using the Born-Oppenheimer approximation. Diffraction-based techniques, such as ultrafast electron and x-ray diffraction offer a unique advantage for imaging the molecular geometry as those measurements are directly sensitive to the spatial distribution of atoms, and are thus complementary to spectroscopic methods that are sensitive to energy differences between electronic states. The nuclear motion in photo-excited molecular crystals has been resolved using ultrafast electron diffraction (UED) with femtosecond resolution[1,2]. Gas phase UED experiments have been successful in capturing the structure of intermediate states with picosecond lifetimes in photoinduced reactions[3–5], in retrieving the three-dimensional structure of transiently aligned molecules[6], and in observing structural deformation due to the interaction with intense laser pulses[7]. Until now, however, it has not been possible to capture the nuclear dynamics in isolated molecules due to the challenges inherent in diffraction experiments from gas samples: The low sample density, the long electron pulse duration due to space charge effects and the velocity mismatch between the probe electrons and the excitation laser pulses.

Laser based time-resolved spectroscopic methods have been used to follow reactions in the gas phase (see e.g. Refs.[8–11]). Their observables are only indirectly related to molecular structure. Recent experiments in gas phase x-ray diffraction using x-ray free electron lasers (XFEL) have shown a sub-100 fs temporal resolution, but the spatial resolution was not sufficient to retrieve atomically resolved structures directly from the data, thus the structures were extracted by a comparison with simulations[12]. In laser-induced electron diffraction (LIED), a high-intensity femtosecond laser pulse is used to ionize a molecule and to then re-scatter the ionized electron from the parent molecule, with resolution of a few femtoseconds[13,14], but so far has not been applied to photoinduced reaction.

One of the limiting factors in gas phase UED has been the velocity mismatch between laser and electron pulses and the temporal broadening of the electron pulses due to Coulomb forces. At megaelectronvolt (MeV) energies the electrons become relativistic, the velocity mismatch is negligible and the Coulomb broadening is significantly reduced. MeV UED has been applied successfully to thin condensed matter samples[15–20], but it was only recently shown that MeV UED can provide sufficient signal to capture the dynamics of molecules in the gas phase. The evolution of a rotational wavepacket in impulsively aligned nitrogen molecules was observed with a resolution of 230 fs full width at half-maximum (FWHM) [21].

Here we show that the coherent motion of a vibrational wavepacket can be experimentally imaged using MeV UED. The method can, in principle, retrieve both the position and the shape of the nuclear wavepackets. In this experiment a vibrational wavepacket on iodine molecules ($I_2$) in the gas phase was excited with a femtosecond laser pulse, and the ensuing motion was observed with sub-angstrom resolution in space and 230 fs resolution in time. The results presented here, combined with previous results on three-dimensional imaging of aligned molecules[6,22], open the door to capturing three-dimensional movies of chemical reactions, where the motion of each nucleus can be observed as the structure evolves from the initial to the final state.

The experimental setup is shown in Fig. 1(a). A vibrational wavepacket is created by resonantly exciting the iodine molecules to the B ($^3\Pi_{0u}$) state with a laser pulse with a central wavelength of 530 nm. The molecules are probed using an electron pulse with 3.7 MeV energy that propagates almost collinearly with the laser pulse through the gas sample. The temporal resolution of the instrument is 230 fs FWHM (100 fs root-mean-square)[21]. See the supplementary materials for a more detailed description of the setup.

Fig.1(b) shows the potential energy surfaces of the ground state and the excited B state of iodine[23]. The figure also shows the amplitude of the ground state wavefunction (blue solid line) and of the excited state wavefunction during the excitation laser (red solid line). The interatomic distance in the excited state oscillates between the Franck-Condon region at 2.7 Å and a maximum separation of 3.9 Å, with a period of approximately 400 fs. Fig. 1 (c) shows a simulation of the coherent dynamics of the wavepacket after excitation, up to a time of 600 fs. Details of the wavepacket simulation are given in the supplemental material.

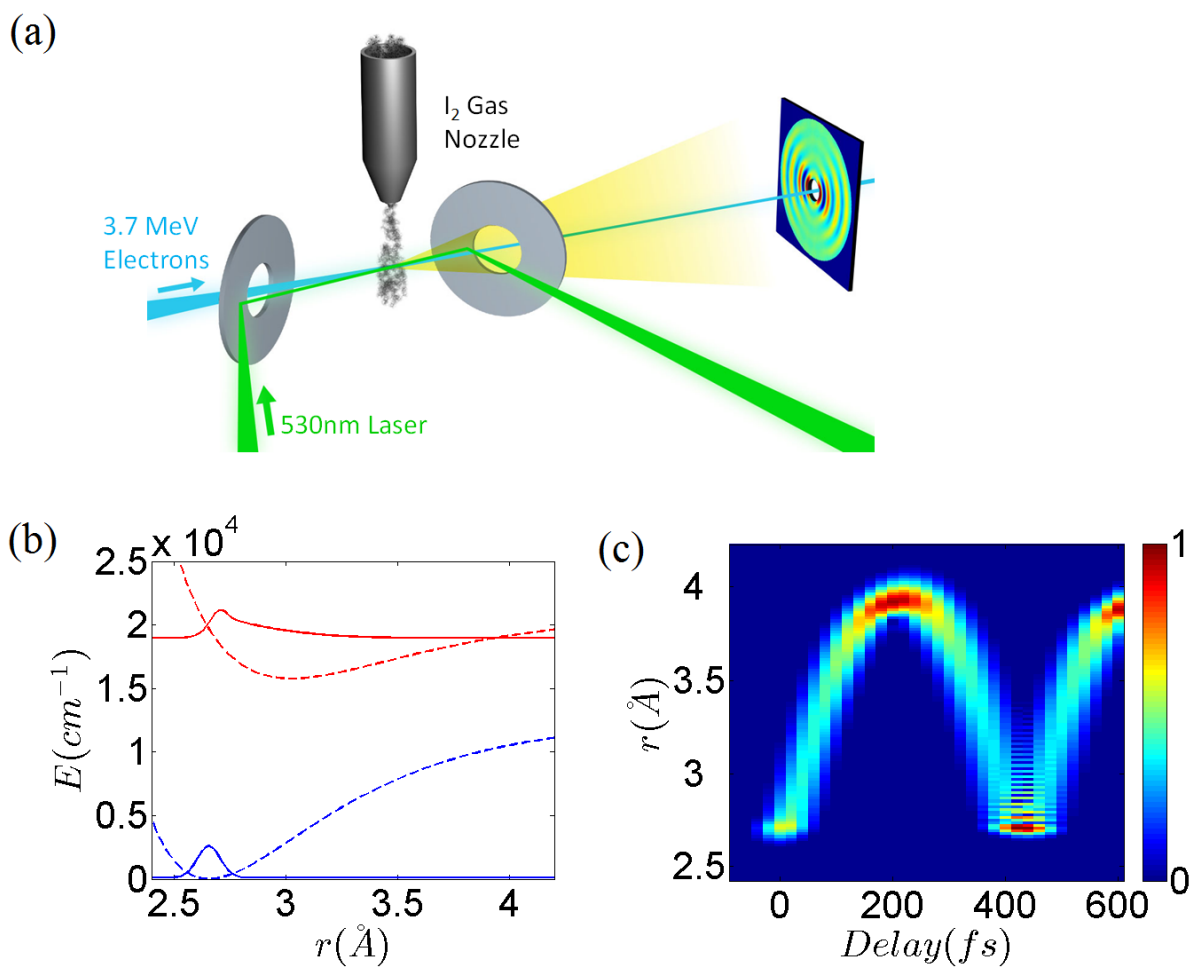

**FIG. 1**. (a) Diagram of the experimental setup showing the laser beam (green), electron beam (blue) and gas jet (gray). (b) Potential energy surfaces of ground state (dashed blue line) and the B excited state (dashed red line), along with the ground and excited state wavepackets (solid lines). The blue and red colors represent ground and the excited states, respectively. (c) Simulated dynamics of the nuclear

wavepacket of $I_2$ in the B state, after excitation by a 530 nm laser pulse. The figure displays the amplitude of the wavefunction as a function of time, in arbitrary units.

We use the standard methods of gas electron diffraction (GED) to extract the molecular structure from the diffraction patterns[26,27]. The diffraction pattern is expressed as a function of momentum transfer $s = \frac{4\pi}{\lambda}\sin(\theta/2)$, where $\lambda$ is the deBroglie wavelength of the electrons, and $\theta$ is the angle between the scattered and transmitted electrons. For the electron energy of 3.7 MeV, $\lambda$ =0.30 pm. Under the independent atom approximation, for randomly oriented molecules the total scattering intensity $I_{tot}$ can be written as the sum of the atomic scattering intensity $I_{at}$ and the molecular scattering intensity $I_{mol}$[28]:

$$I_{at} = \sum_{i=1}^{N}|f_i(s)|^2 \qquad (1)$$

$$I_{mol} = \sum_{i=1}^{N}\sum_{j\neq i}^{N}|f_i(s)||f_j(s)|\cos(\eta_i-\eta_j)\int \frac{\sin(sr)}{sr}P_{ij}(r)dr \qquad (2)$$

where $f_i$, $\eta_i$ are the scattering amplitude and phase of the $i^{th}$ atom, $r$ is the internuclear separation, and $P_{ij}(r)$ is the vibrational probability function for the internuclear distance corresponding to the atom pair $ij$. For diatomic molecules, $P_{ij}(r)$ reduces to the probability density of the nuclear wavefunction, $P(r) = |\chi(r)|^2$. $I_{at}$ can be calculated from the known values of the atomic scattering amplitudes[29]. The structural information of the molecule is usually extracted through the modified diffraction intensity:

$$sM(s) = s\frac{I_{mol}(s)}{I_{at}(s)} \qquad (3)$$

For diatomic molecules such as iodine, using equation (1)-(3), the modified diffraction intensity reduces to

$$sM(s) = \int \frac{\sin(sr)}{r}P(r)dr \qquad (4)$$

In the case of static structures, $P(r)$ is a very narrow distribution centered at the equilibrium distance, and the interatomic distance can be directly extracted from the period of the sinusoidal modulation in $sM(s)$. The experimental $sM(s)$ from the static $I_2$ diffraction patterns is shown in the supplementary material. The nuclear wavefunction probability $P(r)$ can be calculated by a sine transform of $sM$:

$$P(r) = r \cdot \int_0^{s_{Max}} sM(s)\sin(rs)e^{(-ks^2)}ds \qquad (5)$$

where $s_{Max}$ is the maximum momentum change measured in the diffraction pattern, and $k$ is a damping constant that is used to reduce artifacts in the transform.

In order to highlight the changes in the diffraction patterns, we use the diffraction-difference method[3,30]. This method compares diffraction patterns before and after laser excitation. A difference diffraction intensity map is calculated: $\Delta I(t,s,\phi) = I(t,s,\phi) - I(-T,s)$, where $t$ is some time after laser excitation, $\phi$ is the azimuthal angle in the diffraction pattern, and $-T$ represents a negative time, i.e. before the molecules are excited by the laser. Molecules that lie along the direction of the laser polarization are more likely to be excited, with the probability of excitation having a $\cos^2(\alpha)$ dependence, where $\alpha$ is the angle between the molecular axis and the laser polarization. Thus, diffraction patterns after excitation become anisotropic, which has been previously observed with electron diffraction in $C_2F_4I_2$ molecules[31], where the lifetime was found to be approximately 3 ps. Fig. 2(a) shows the experimental pattern $\langle \Delta I_E \rangle$, which is averaged over multiple diffraction patterns at time delays between 50 fs and 550 fs (roughly one vibrational period). The pattern shows clear evidence of changes in the interatomic distance in the diffraction rings, as well as a significant anisotropy. Fig. 2(b) shows the theoretical pattern $\langle \Delta I_S \rangle$ that was simulated using the wavepacket shown in Fig. 1(c) time-averaged from 50 fs to 550 fs, and assuming a $\cos^2(\alpha)$ dependence on the excitation probability. In the simulation resulting in Fig. 2 (b) the diffraction pattern was calculated using the formalism from Ref. [32], taking into account the angular distribution resulting from the excitation process. The theory matches all the changes in the diffraction rings and the anisotropy of the experimental pattern closely. We have also monitored the anisotropy in the diffraction pattern over a longer time window and seen that it decays with a time constant of 1.5 ps, longer than the time window over which the vibrational motion is observed. In Fig. 2(a) and 2(b), the laser polarization direction is horizontal.

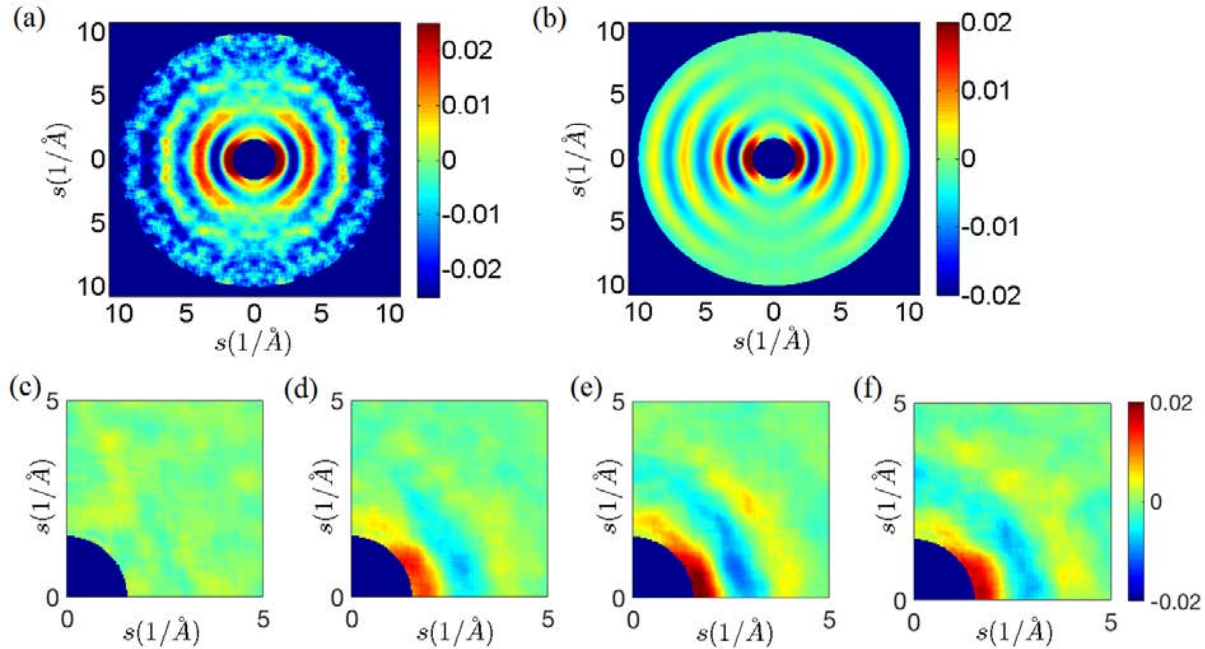

**FIG. 2** (a) Experimental difference diffraction pattern $\langle \Delta I_E(t,s,\phi) \rangle$ averaged over $t = 50$ fs to 550 fs. (b) Simulated difference diffraction pattern $\langle \Delta I_{sim}(t,s,\phi) \rangle$ averaged over $t = 50$ fs to 550 fs. The color scales are in arbitrary units. (c) through (f): Experimental difference diffraction patterns $\Delta I_E(t,s,\phi)$ at $t = -184$ fs, $t = 17$ fs, $t = 217$ fs and $t = 417$ fs, respectively.

Figure 2(c)-(f) shows the experimental difference pattern $\Delta I_E(t,s,\phi)$ at four consecutive time delays with a time step of 200 fs. The first pattern is before time zero, while the second is shortly after time-zero. The patterns show clear changes in the diffraction pattern. For each time, an azimuthal average is used to reduce the data to 1-D. At the low level of alignment generated by excitation photo-selection, the extraction of the interatomic distance is not significantly affected by using the azimuthally averaged data, or by changes in the angular distribution that occur within the time window of the experiment. The modified difference intensity is defined as:

$$\Delta sM(t,s) = s \frac{\langle \Delta I(t,s,\phi) \rangle}{I_{at}(s)}, \qquad (6)$$

where the averaging is over the azimuthal angle $\phi$.

Fig. 3 (a) and (b) shows the experimental $\Delta sM_{exp}(t,s)$ and the simulated $\Delta sM_{sim}(t,s)$ for pump-probe delays between $t = -300$ fs and $t = 550$ fs, with a step size $\Delta t$ of 67 fs. The patterns at negative time delay correspond to the probing electrons arriving at the sample before the laser pulse. In the simulation, the effect of the temporal resolution is included by a convolution with a Gaussian kernel possessing a FWHM of 230 fs. The total exposure time for each time delay is approximately 15 minutes. Fig. 3 (a) displays $\Delta sM_{exp}(t,s)$ over an s-range of $1.6$ Å$^{-1}$ $< s <10$ Å$^{-1}$. Fig. 3 (b) shows the simulated $\Delta sM_{sim}(t,s)$ over a similar range but including the information at low scattering angles not accessible experimentally. Experiment and simulation are in good agreement, with the experimental results becoming noisier for the larger $s$ values where there are fewer counts but still showing the same trend as in the theory. The interatomic distance as a function of time can be extracted directly from modified molecular scattering of the excited state $sM_{Excited}(t,s)$. $\Delta sM(t,s)$ can be expressed as the difference between the excited and not excited molecules, multiplied by the excitation factor:

$$\Delta sM(t,s) = \epsilon\, (sM_{Excited}(t,s) - sM_{Ground}(s)) \qquad (7)$$

where $sM_{Excited}$ and $sM_{Ground}(s)$ are the modified scattering intensities of the ground and excited states and $\epsilon$ the fraction of optically excited molecules. Comparing the simulated $\Delta sM(t,s)$ to the experimental values allowed us to deduce an excitation fraction of 15%.

We reconstruct $sM_{Excited}(t,s)$ from the difference $\Delta sM(t,s)$ by adding the known contribution from ground state, taking into account the excitation factor $\epsilon$. The ground state contribution is not time dependent, and was calculated using equation (3). Fig. 3(c) and 3(d) show the experimental and the simulated $sM_{Excited}(t,s)$, repectively. Again there is good agreement between theory and experiment. In this case, since the patterns reflect only the contribution from the excited molecules, the most probable interatomic distance $r(t)$ can be extracted directly from the period of the interference pattern (see equation 4).

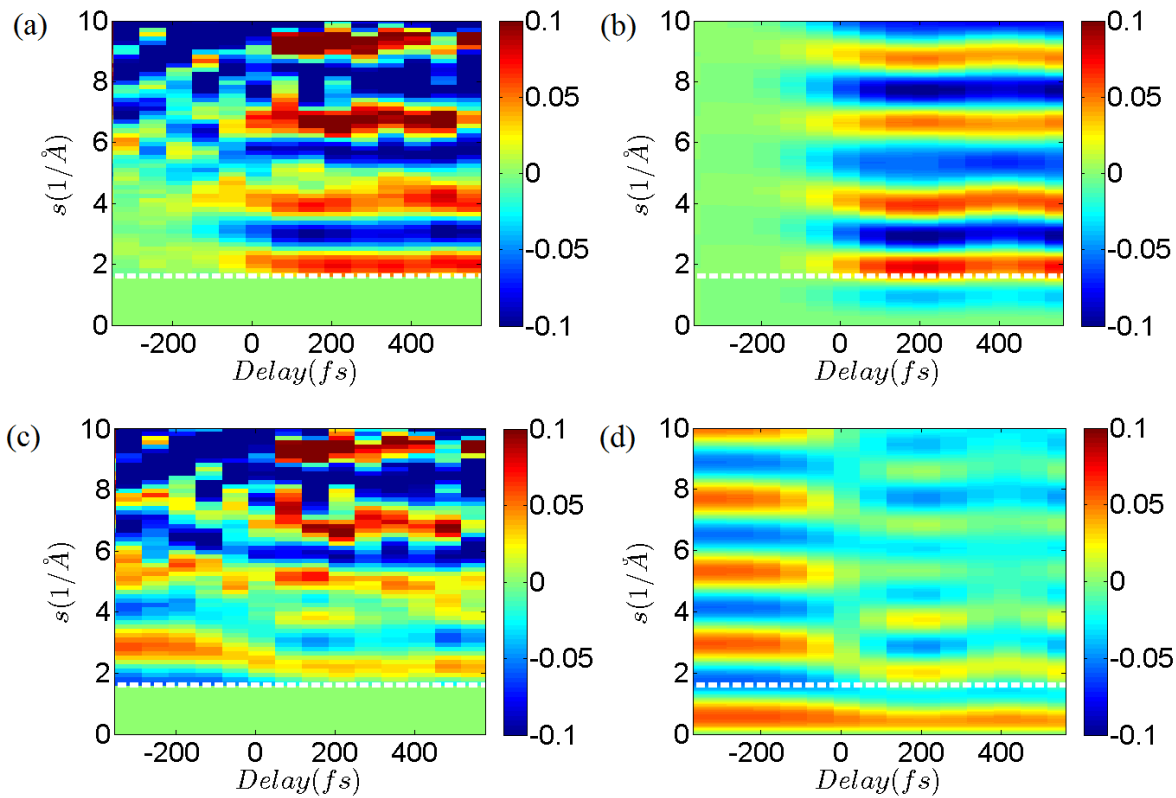

**FIG. 3**. (a,b) The experimental (a) and simulated (b) time-resolved modified scattering intensity difference $\Delta sM(t,s)$. (c, d) The experimental (c) and simulated (d) time-resolved modified scattering intensity of the excited state $sM_{Excited}(t,s)$. The pattern in (c) was generated by adding the known contribution from ground state to the experimental pattern in (a). The pattern in (d) was generated by adding the known contribution from ground state to the simulated pattern in (b). The experiment misses $s$ values between 0 and 1.6 Å$^{-1}$ (area underneath the white dashed line in each pattern) due to a hole in the detector that serves to transmit the main electron beam.

Finally, we extract the bond length $r(t)$ by fitting a sine function to the experimental $sM_{Excited}(t,s)$, using a least-square fitting routine[26,27]. The fit was performed over an $s$ range of 1.6 Å$^{-1}$ to 9.7 Å$^{-1}$, and the data was weighted by the signal to noise ratio (SNR). Fig. 4 (a) shows the bond lengths extracted from the data, which clearly show the first 1.5 periods of the wavepacket motion. The error bars represent the standard errors from the least-square fitting for each point. The precision of the measurement can be estimated from the average value of the standard errors, which is 0.07 Å. The same fitting procedure was used also to extract the bond length in simulated diffraction patterns. The dashed red line shows the results for *r(t)* from the simulation, which is in close agreement with the experiment.

A key feature of diffraction is that it is sensitive not only to the distance between the atoms, but also to the probability density in the nuclear wavefunction $P(t,r)$. Fig. 4(b) shows the experimentally retrieved $P(t,r)$ as a function of time (blue lines). The experimental $P(t,r)$ was obtained by applying equation (5) (with a value of k = 0.05) at each time slice of the data in Fig. 3(c), after filling in the missing *s*<1.6 Å$^{-1}$ area with values obtained from the fitting routine used to generate Fig. 4(a). The width of $P(t,r)$ before time-zero is 0.7 Å FWHM, which is determined by the spatial resolution of the experiment. Around t = 0, the wavefunction starts to extend towards larger interatomic distances and becomes asymmetric. During the motion of the nuclei, the width of $P(r)$ increases to between 1.0 Å and 1.3 Å. The spread is caused by the motion of the wavepacket, averaged over the temporal resolution of the experiment. The measured results are in good agreement with the theoretical $P(t,r)$ calculated from the theoretical *sM* in Fig. 3(d). The dashed red lines in Fig. 4 (b) are the theoretical calculations assuming the same spatial resolution as that of the experiment, while the solid green lines include also the effect of the temporal resolution (230 fs FWHM). Both the experimental and the theoretical $P(t,r)$ are normalized to the peak of the top curve $P(-133\,fs,r)$. While in this case the measured broadening is mostly due to time-averaging of the motion, with improved spatial and temporal resolution we expect that the intrinsic changes in the shape of the wavepacket will become accessible. For example, the calculations shown in Fig. 1 (c) show that initially after excitation, the width of the wavepacket is 0.1 Å, and the width increases to 0.7 Å at the midpoint between the inner and outer turning points of the motion. These effects would become visible with a factor-of-two improvement in the spatial resolution and a factor-of-three improvement in temporal resolution. Thus, the method is suitable to read fine details of the wavepacket and distinguish coherent states from squeezed states.

In conclusion, we have used MeV UED to image the motion of a vibrational wavepacket in iodine. With the spatiotemporal resolution of the current setup, many interesting photochemical reactions are already within reach. For example, the photoisomerization of azobenzene molecules which proceeds with a time constant of 420 fs[33] and relaxation dynamics during $CS_2$ dissociation, which include a periodic motion with a time constant of ~900 fs [10,34,35] Methods recently developed for three-dimensional molecular imaging with UED [6,22] can be combined with the femtosecond capability demonstrated here to capture three-dimensional movies of these and other molecular reactions.

With improved spatial and temporal resolution, it will be possible to capture also the spreading of nuclear wavepackets in molecules. This is of crucial importance in the context of excited state chemical dynamics. In most cases, the wavepacket does not stay nicely focused but spreads due to the anharmonic nature of the potential energy surfaces. In addition, anharmonic coupling spreads the wavepacket density over many normal modes. Additional improvements in spatiotemporal resolution can be achieved by using RF compression[36,37] to reduce the pulse duration and to increase the number of electrons per pulse. An alternative approach is to operate the experiment at higher repetition rate using less charge per pulse to reduce the temporal broadening due to Coulomb forces. Further improvements of the temporal resolution to sub-50 fs will require counteracting the effect of timing jitter, either by actively compensating it or by measuring the relative time of arrival of each pulse[38].

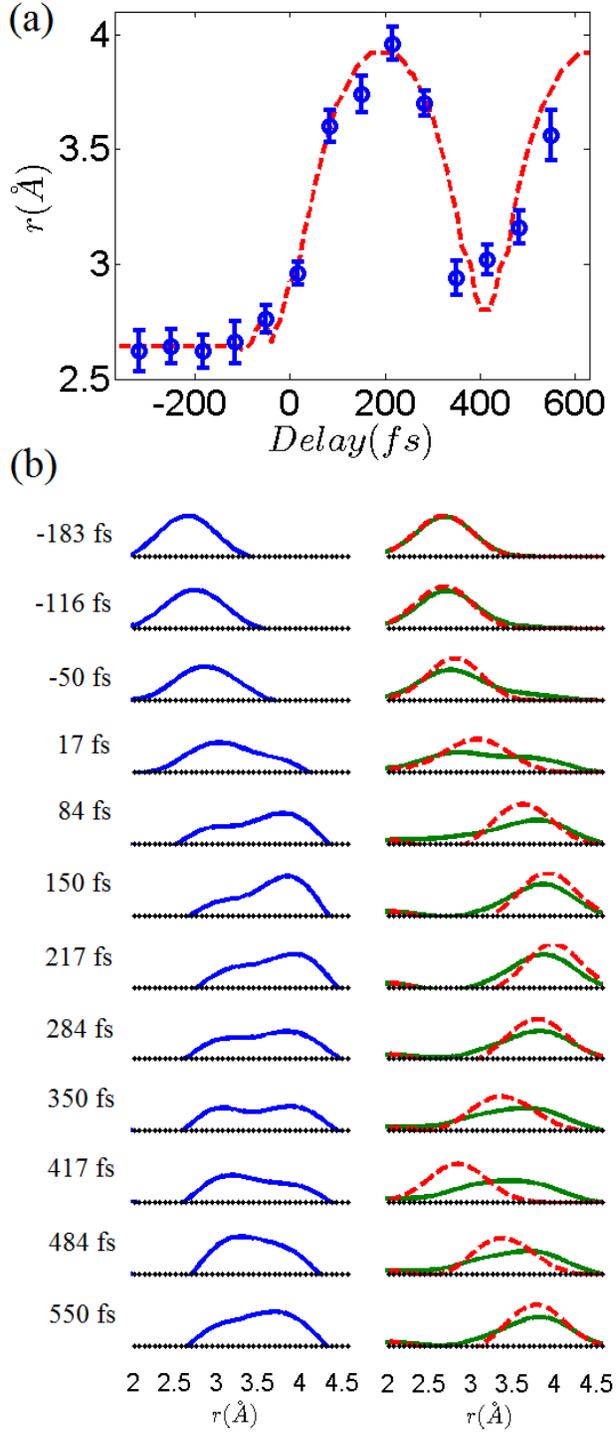

**FIG.** 4.(a) Time-resolved experimental (blue circles with error bars) and simulated (red dashed line) bond lengths from a single sine fit of sM. (b) The experimental (left panel) and simulated (right panel) time-resolved probability density function of the wavepacket $P(t,r)$, calculated using equation (5). The dashed red line is the results of the theoretical calculation with the same

spatial resolution as the experiment, and the solid green line includes also averaging due to the temporal resolution (230 fs). The black dotted line indicates the baseline for each curve.

**Acknowledgements**

The authors would like to thank SLAC management for the strong support. The technical support by SLAC Accelerator Directorate, Technology Innovation Directorate, LCLS Laser Science & Technology Division and Test Facilities Department is gratefully acknowledged. This work was supported in part by the U.S. Department of Energy (DOE) Contract No. DE-AC02-76SF00515, DOE Office of Basic Energy Sciences Scientific User Facilities Division, the SLAC UED/UEM Initiative Program Development Fund, and by the AMOS program within the Chemical Sciences, Geosciences, and Biosciences Division of the Office of Basic Energy Sciences, Office of Science, U.S. Department of Energy. J. Yang, and M. Centurion were partially supported by the U.S. Department of Energy Office of Science, Office of Basic Energy Sciences under Award Number DE-SC0014170. MG is now supported by a Lichtenberg Professorship from the Volkswagen foundation. M. S. Robinson was supported by the National Science Foundation EPSCoR RII Track-2 CA Award No. IIA-1430519.

**Author contributions**

J.Y., M.G., T.V., M. S. R., R.L., X.S., S.W, and X.W. carried out the experiments. N.H., R. C., J. C., I. M., S. V., and A.F. developed the laser system. M.G. and J.Y. constructed the setup for gas phase experiments. C. H., K. J., A. R., and C. Y. helped on experimental setup. J.Y. and M. C. performed the data analysis. M.G. and J. Y. performed the simulations. The experiment was conceived by M.G., M.C. and X.W. The manuscript was prepared by J.Y. and M.C. with discussion and improvements from all authors. M.C. and X.W. supervised the work.